\documentclass[usenatbib]{emulateapj}
\usepackage{graphicx,amsmath,amssymb,hyperref,xspace,mathrsfs}

\newcommand{\D}{\mathrm{d}}
\renewcommand{\d}{\partial}
\newcommand{\Bx}{\boldsymbol{x}}
\newcommand{\Bv}{\boldsymbol{v}}
\newcommand{\rmax}{r_\mathrm{max}}
\newcommand{\Rlc}{\mathcal{R}_\mathrm{LC}}
\newcommand{\scE}{\mathscr E}
\newcommand{\hmin}{h_\mathrm{min}}
\newcommand{\hinfl}{h_\mathrm{infl}}
\newcommand{\fold}{\boldsymbol f^\mathrm{old}}
\newcommand{\fnew}{\boldsymbol f^\mathrm{new}}

\begin{document}

\title[Fokker--Planck models]
{A new Fokker--Planck approach for relaxation-driven evolution of galactic nuclei}
\author{Eugene Vasiliev\altaffilmark{1,2}}
\email{eugvas@lpi.ru}
\affil{$^{1}$Rudolf Peierls Centre for Theoretical Physics, 1 Keble road, Oxford, UK, OX1 3NP}
\affil{$^{2}$Lebedev Physical Institute, Leninsky prospekt 53, Moscow, Russia, 119991}

\begin{abstract}
We present an approach for simulating the collisional evolution of spherical isotropic stellar systems based on the one-dimensional Fokker--Planck equation. A novel aspect is that we use the phase volume as the argument of the distribution function, instead of the traditionally used energy, which facilitates the solution. The publicly available code, \textsc{PhaseFlow}, implements a high-accuracy finite-element method for the Fokker--Planck equation, and can handle multiple-component systems, optionally with the central black hole and taking into account loss-cone effects and star formation. \protect\\
We discuss the energy balance in the general setting, and in application to the Bahcall--Wolf cusp around a central black hole, for which we derive a perturbative solution. We stress that the cusp is not a steady-state structure, but rather evolves in amplitude while retaining an approximately $\rho\propto r^{-7/4}$ density profile. \protect\\
Finally, we apply the method to the nuclear star cluster of the Milky Way, and illustrate a possible evolutionary scenario in which a two-component system of lighter main-sequence stars and stellar-mass black holes develops a Bahcall--Wolf cusp in the heavier component and a weaker $\rho\propto r^{-3/2}$ cusp in the lighter, visible component, over the period of several Gyr. The present-day density profile is consistent with the recently detected mild cusp inside the central parsec, and is weakly sensitive to initial conditions. 
\end{abstract}
\maketitle

\section{Introduction}  \label{sec:Introduction}

The dynamical evolution of many classes of stellar systems is driven by two-body (collisional) relaxation. While a rigorous theoretical description of this phenomenon needs to take into account spatial inhomogeneities and gravitational polarization effects \citep[e.g.,][]{Heyvaerts2010, Chavanis2012}, this leads to extremely complicated equations, and in practice simpler approaches are usually taken. One of the commonly used approximations is that the overall effect of two-body relaxation may be described as a sequence of uncorrelated pairwise weak encounters of a test star moving through a uniform infinite medium of field stars. The evolution of the distribution function (DF) of these test stars is described in terms of a Fokker--Planck equation, with advection (drift) and diffusion coefficients for velocity computed from the DF of the field stars \citep{Rosenbluth1957}.
Furthermore, recognizing that the cumulative effect of perturbations is small over the dynamical timescale, the Fokker--Planck equation is written in the orbit-averaged way, for the DF expressed in terms of integrals of motion. Finally, identifying the DF of test stars with that of the field stars and mandating that the gravitational potential is determined by the density computed from the DF itself, one arrives at the coupled system of Fokker--Planck and Poisson equations describing the evolution of a stellar system under two-body relaxation. This was first accomplished by \citet{Henon1961} for a homologous spherical isotropic model, and later by \citet{Cohn1979} for a spherical system with a DF $f$ expressed in terms of energy $E$ and angular momentum $L$ on a suitable grid in this two-dimensional space. In his approach the DF of field stars, entering the expressions for diffusion coefficients, was taken to be the isotropized version of $f(E,L)$, i.e., a function of $E$ only. This approximation greatly simplifies the calculations and is justified by the fact that the diffusion coefficients are integrals over the DF, hence should be relatively insensitive to its moderate deviation from isotropy. Furthermore, the solution itself was found to depend on $L$ rather weakly. Recognizing this fact, \citet{Cohn1980} simplified the equations even further by considering the isotropic one-dimensional Fokker--Planck equation for $f(E)$, which substantially increased numerical accuracy.

The next decade has seen great progress in applying this approach to the dynamics of star clusters and galactic nuclei. Several groups have developed independent implementations of the one-dimensional isotropic Fokker--Planck equation, adding various layers of complexity: two or more mass components \citep{InagakiWiyanto1984}, strong scattering \citep{Goodman1983}, stellar mergers and three-body heating \citep{Lee1987, QuinlanShapiro1989}, stellar mass loss and tidal escape \citep{ChernoffWeinberg1990}, the loss of stars into the central black hole (\citealt{Murphy1991}, following the earlier work of \citealt{CohnKulsrud1978}). Later, new methods for solving the two-dimensional equation were presented in \citet{Takahashi1995,Drukier1999,EinselSpurzem1999}. One-dimensional approximation is sufficient for many problems of interest, and agrees with the more general two-dimensional treatment reasonably well \citep{Cohn1985}.
At the same time alternative approaches were developed: spherical Monte Carlo \citep[e.g.,][]{Shapiro1985} and gaseous \citep[e.g.,][]{LouisSpurzem1991} models, and a large industry of $N$-body simulations. Fokker--Planck models still enjoy some popularity, and owing to their low computational demand, allow to explore quickly a large parameter space. So far none of the existing codes were made public, which certainly limits their usage; we intend to fill this gap with a new implementation which is described in this paper.

The novel aspect of our method is the use of phase volume as the argument of DF, which is a one-dimensional projection of the three-dimensional action space. The use of actions in Fokker--Planck calculations is conceptually cleaner \citep[e.g.,][]{BinneyLacey1988} and simplifies the recomputation of gravitational potential (the DF evolves adiabatically and hence is unchanged when expressed in terms of actions). Nevertheless, for unclear reasons it has never been attempted, except for the unpublished thesis by \citet{Girash2009}. Another novel feature of our implementation is the use of finite-element method for the Fokker--Planck equation; with higher-order polynomial basis elements this substantially increases the accuracy of spatial discretization compared to the traditionally employed \citet{ChangCooper1970} finite-difference scheme. A similar approach was used in \citet{Takahashi1993}, although it was not labelled as such.

We present the basic formalism for dealing with phase volume and the moments of DF in Section~\ref{sec:Definitions}, introduce the Fokker--Planck equation with all auxiliary ingredients in Section~\ref{sec:FokkerPlanck}, and discuss its conservation properties in Section~\ref{sec:Conservation}. Then we revisit the classical problem of the cusp formation around a central massive black hole in Section~\ref{sec:BWcusp}, and derive the first-order correction to the scale-free Bahcall--Wolf solution, demonstrating that it does not remain stationary, but rather changes its amplitude while approximately retaining the functional form. In Section~\ref{sec:Validation} we validate our Fokker--Planck code against more sophisticated modelling methods in a test case of a re-growing cusp. Finally, we apply our code to the nuclear star cluster of our Galaxy in Section~\ref{sec:MilkyWayNSC}, and demonstrate that the present-day structure agrees with the observations under quite general assumptions. Section~\ref{sec:Summary} sums up. Technical details are deferred to the \hyperlink{sec:Appendix}{Appendix}.

\section{Definitions}  \label{sec:Definitions}

We consider a spherically-symmetric stellar system, consisting of one or several species of stars (components), each one described by a distribution function (DF) $f_c(\Bx,\Bv)$, $c=1..N_\mathrm{comp}$. We use the convention that the integral $\iiint \D^3x \iiint \D^3v\; f_c(\Bx,\Bv) = M_c$, the total mass of stars in this component. The mass of an individual star of each species is denoted as $m_c$.
The total gravitational potential is denoted by $\Phi(r)$, and its inverse function is $\rmax(E)$, the maximum radius accessible to a star with energy $E$ (so that $\Phi(\rmax(E))=E$).

According to the Jeans' theorem, a steady-state DF $f(\Bx,\Bv)$ must be a function of integrals of motion -- in a spherical system, these are the energy $E \equiv \Phi(|\Bx|) + \frac12|\Bv|^2$ and the angular momentum $L \equiv |\Bx\times\Bv|$; in the isotropic case, $f$ may only depend on $E$.

Instead of $E$, we use the phase volume $h$ as the argument of the DF.
It is defined as the volume of phase space enclosed by the energy hypersurface:
\begin{subequations}
\begin{align}
h(E) &\equiv \iiint \D^3 x \iiint \D^3 v\; \left\{  \begin{array}{ll}
\! 1 &\mbox{if }\, \Phi(|\Bx|) + |\Bv|^2/2 < E, \\
\! 0 &\mbox{otherwise} \end{array} \right.  \nonumber\\
&= \int_0^{\rmax(E)} 4\pi\,r^2\,\D r \int_0^{\sqrt{2[E-\Phi(r)]}} 4\pi\, v^2\,\D v  
\end{align}
\begin{align}
&= \frac{16\pi^2}{3} \int_0^{\rmax(E)}
   r^2\, \Big\{ 2 \big[E-\Phi(r)\big]\Big \}^{3/2}\; \D r  \label{eq:h_int_r} \\
&= 16\pi^2 \int_{\Phi(0)}^{E}
   \D E'\, \int_0^{\rmax(E')} r^2 \sqrt{ 2 \big[E'-\Phi(r)\big] }\; \D r  \label{eq:h_int_g}\\
&= 4\pi^2 \int_0^{L^2_\mathrm{circ}(E)} J_r(E,L)\; \D L^2 .  \label{eq:h_int_Jr}
\end{align}
\end{subequations}
Its derivative by energy is called the density of states%
\footnote{\citet{Cohn1980} and other studies use two similarly related quantities: $p\equiv g/(4\pi^2), q\equiv h/(4\pi^2)$.}:
\begin{subequations}
\begin{align}
g(E) &\equiv \frac{\D h(E)}{\D E}  \nonumber\\
     &= 16\pi^2 \int_0^{\rmax(E)} r^2\, \sqrt{ 2 \big[E-\Phi(r)\big] }\; \D r  \label{eq:g_int_r}\\
     &= 4\pi^2\, \int_0^{L_\mathrm{circ}^2(E)} T_\mathrm{rad}(E,L)\; \D L^2 .
\end{align}
\end{subequations}

Here
$J_r(E,L) \equiv \int_{r_-}^{r_+} v_r\, \D r$ is the radial action,
$T_\mathrm{rad}(E,L) \equiv 2 \int_{r_-}^{r_+} \D r/v_r = \d J_r/\d E$  is the radial period (its dependence on $L$ at a fixed $E$ is usually weak),
and $L_\mathrm{circ}(E)$ is the angular momentum of a circular orbit with energy $E$.
We note that in the case of a two-dimensional anisotropic Fokker--Planck equation one could use two action variables $J_r$ and $L$ as arguments of the DF; $h$ is the counterpart (\ref{eq:h_int_Jr}) of these action variables in the one-dimensional case. \citet{Cohn1979} and later studies did express the DF in terms of $J_r$ and $L$ during the recomputation of potential, but still used $f(E,L)$ in the Fokker--Planck equation itself.

The correspondence between energy $E$ and phase volume $h$ is determined by the potential $\Phi(r)$;
since both $h$ and $g$ are monotonically increasing functions of $E$, this is an invertible transformation, so one may equivalently express $E(h)$ and $g(h)$.
Conversely, given $E(h)$, one may find the potential $\Phi(r)$ (or, rather, the inverse function $\rmax(\Phi)$) using the Abel transform:
\begin{align}
\rmax^3(\Phi)
&= \frac{3}{8\pi^3} \int_{\Phi(0)}^\Phi
   \frac{\D E}{\sqrt{2(\Phi-E)}} \frac{\D g(E)}{\D E}  \nonumber\\
&= \frac{3}{8\pi^3} \int_0^{h(\Phi)}
   \frac{\D h}{\sqrt{2[\Phi-E(h)]}} \frac{\D g(h)}{\D h} .
\end{align}

In a power-law potential $\Phi(r) = \Phi_0 + C r^{2-\gamma}$ (which corresponds to a density profile $\rho\propto r^{-\gamma}$), the functions $g$ and $h$ also have a power-law behaviour:
$g(E) \propto \mathcal{E}^{(8-\gamma)/(4-2\gamma)}$, $h(E) \propto \mathcal{E}^{(12-3\gamma)/(4-2\gamma)}$, where $\mathcal E \equiv E-\Phi_0$ if the potential is finite at origin ($\gamma<2$), or $\mathcal E \equiv -E$ otherwise. In both cases, $g(h) \propto h^{(8-\gamma)/(12-3\gamma)}$. In particular, for a Kepler potential $\Phi=-GM/r$ (either in the vicinity of the central black hole, or at large radii where the density is negligible), 
\begin{align}  \label{eq:Eofh_Kepler}
h = \frac{2\sqrt{2}\,\pi^3\, (GM)^3}{3\, (-E)^{3/2}}, \quad
g = \frac{(3h)^{5/3}} {4\pi^2\,(GM)^2} = -\frac32 \frac{h}{E}.
\end{align}

For a given DF $f(h)$, one may introduce several derived functions of phase volume:
\begin{align}
I_0(h) &\equiv \int_{E(h)}^0 f(E')\, \D E'
      &&= \int_h^\infty \frac{f(h')}{g(h')}\, \D h' , \!\!\!\label{eq:I_0} \\
\!\!K_g(h) &\equiv \int_{E(0)}^{E(h)} f(E')\,g(E')\, \D E'
      &&= \int_0^h f(h')\, \D h' , \!\!\!\label{eq:K_g} \\
\!\!K_h(h) &\equiv \int_{E(0)}^{E(h)} f(E')\,h(E')\, \D E'
      &&= \int_0^h \frac{f(h')\,h'}{g(h')}\, \D h' , \!\!\!\label{eq:K_h} \\
\!\!K_E(h) &\equiv \int_{E(0)}^{E(h)}\! f(E')\,g(E')\, E'\, \D E' \!\!\!\!\!\!
      &&= \int_0^h\! f(h')\,E(h')\, \D h' .\;\;\;{}\nonumber
\end{align}
It is easy to demonstrate that
$K_g(h)$ is the mass of stars with energies less than $E$ (or enclosed by phase volume $h$),
$K_h(h)$ equals to $2/3$ times the kinetic energy of stars enclosed by phase volume $h$, and
$K_E(h)$ measures the total energy (sum of kinetic and potential energies) of stars inside this volume. 
For instance, writing the kinetic energy as
$\iiint \D^3x \iiint \D^3v\;f(\Bx,\Bv)\,|\Bv|^2/2$ and transforming the integration volume to
$\int_{\Phi(0)}^E \D E'\, \int_0^{\rmax(E')} \D r$, as in (\ref{eq:h_int_g}), one obtains (\ref{eq:K_h}).

A physically valid system must have finite mass, thus $f(h)$ must drop faster than $h^{-1}$ as $h\to\infty$, and rise slower than $h^{-1}$ as $h\to0$. Additionally, the requirement for the energy to be finite imposes a stricter constraint for the inner DF slope in the case of a singular potential ($\gamma\ge 2$): $f(h)$ should grow no faster than $h^{(5\gamma-16)/(12-3\gamma)}$; in the Kepler case ($\gamma=3$) this reads $f \lesssim C h^{-1/3} \propto \sqrt{-E}$, which implies that the density profile must be shallower than $\rho\propto r^{-2}$.

The advantage of using $h$ as the argument of the DF is that it is conserved under adiabatic changes of the potential \citep[e.g.,][]{Young1980}, which will be important for the Fokker--Planck equation. The density is related to the DF via
\begin{align}  \label{eq:rho_from_f}
\rho(r) &= 4\pi \int_{\Phi(r)}^{0} \D E\;f(E)\; \sqrt{2\big[E-\Phi(r)\big]} \\
&= 4\pi \int_{h[\Phi(r)]}^\infty \D h'\; \frac{f(h')}{g(h')}\;
\sqrt{2\big[E(h')-\Phi(r)\big]} ,\nonumber
\end{align}
and determines the potential through the Poisson equation, which in the spherically-symmetric case yields
\begin{align}  \label{eq:Poisson}
\Phi(r) = -4\pi G \left[ \frac{1}{r} \int_0^r \D r'\; r'^2\;\rho(r') +
\int_r^\infty \D r'\; r'\;\rho(r') \right].
\end{align}

\section{Fokker--Planck equation}  \label{sec:FokkerPlanck}

The one-dimensional orbit-averaged Fokker--Planck equation describing the diffusion in energy space can be written in the following flux-conservative form \citep[e.g.,][]{Cohn1980}:
\begin{subequations}
\begin{align}
\frac{\d\, [\,f(E,t)\,g(E)\,]}{\d t} &= - \frac{\d \mathcal F(E,t)}{\d E}, \\
-\mathcal F(E,t)  &\equiv D_{EE}(E)\, \frac{\d f(E,t)}{\d E} + D_E(E)\,f(E,t) .
\end{align}
\end{subequations}
We take $h$ as the independent variable instead of $E$, transforming the derivatives as 
$\frac{\d}{\d E} = g(h)\,\frac{\d}{\d h}$. We also add source $s$ and sink $-\nu f$ terms to this equation, and write it separately for each species $c$:
\begin{subequations}  \label{eq:FokkerPlanck}
\begin{align}
\frac{\d f_c(h,t)}{\d t} &= - \frac{\d \mathcal F_c(h,t)}{\d h} + s_c(h,t) - \nu_c(h,t)\, f_c(h,t),
\label{eq:FokkerPlanckDt} \\
-\mathcal F_c  &\equiv A_c\,f_c + D\, \frac{\d f_c}{\d h}.  \label{eq:MassFlux}
\end{align}
\end{subequations}
Here $\mathcal F$ is the flux through the phase volume, and the advection and diffusion coefficients are given by
\begin{subequations}  \label{eq:FokkerPlanckCoefs}
\begin{align}
A_c(h) &= \Gamma\; m_c\;\; \sum_i K_{g,i}(h) ,  \label{eq:D_h} \\
D  (h) &= \Gamma\, g(h)\,  \sum_i m_i\,\big[ h\, I_{0,i}(h) + K_{h,i}(h) \big] ,  \label{eq:D_hh} \\
\Gamma     &\equiv 16\pi^2\,G^2\,\ln\Lambda,
\end{align}
\end{subequations}
where $\ln\Lambda$ is the Coulomb logarithm, and the functions $I_{0,i}, K_{g,i}$ and $K_{h,i}$ are given by (\ref{eq:I_0}-\ref{eq:K_h}) for each species. Note that the diffusion coefficient $D$ is the same for all species, while the advection coefficient $A_c$ is proportional to the mass of a single star for each species.

The source term $s_c(h)$ may represent the star formation rate per unit phase volume; it is not localized in the real space, but rather spread according to the density generated by a function of $h$ (\ref{eq:rho_from_f}) -- in this case, the rate of increase of $f(h)$ with time. A $\delta$-function source at $h_0$ corresponds to the star formation rate $\dot\rho(r) \propto \sqrt{E(h_0)-\Phi(r)}$, i.e., somewhat concentrated towards the origin.

The sink term may describe the loss of stars captured by the central black hole. Of course, the capture or tidal disruption occurs when the star passes the pericenter of its orbit at a distance less than $r_\mathrm{LC}$ from the black hole, so any description in terms of orbit-averaged Fokker--Planck equation is necessarily approximate. By returning to the local (non-orbit-averaged) equation, \citet{CohnKulsrud1978} derived a suitable boundary condition for a one-dimensional orbit-averaged diffusion equation in angular momentum, neglecting the diffusion in energy. It is expressed in terms of the loss-cone filling factor $q$, the ratio between the mean-square change of angular momentum per one orbital period to the width of the loss-cone boundary:
\begin{subequations}
\begin{align}
q(E) &\equiv \frac{\mu\;T_\mathrm{rad}(E,L=0)}{\Rlc} ,  \label{eq:q_losscone} \\
\Rlc(E) &\equiv \frac{L^2_\mathrm{LC}(E)}{L^2_\mathrm{circ}(E)} =
\frac{2\,G\,M_\bullet\,r_\mathrm{LC}}{ L^2_\mathrm{circ}(E) } , \\
\mu(E) &\equiv \frac{8\pi^2}{g(E)}  \int_0^{r_\mathrm{max}(E)}
\frac{\langle \Delta v_\bot^2 \rangle\; r^2\,\D r}{\sqrt{2\big[E-\Phi(r)\big]}} , \\
\langle \Delta v_\bot^2 \rangle &\equiv \sum_i \Gamma m_i
\bigg[ \frac{4}{3}I_{0,i}(E) + 2 J_{1/2,i} - \frac{2}{3} J_{3/2,i} \bigg] ,
\label{eq:DeltaV2per} \\
J_n(E,r) &\equiv \int_{\Phi(r)}^E \D E'\;f_i(E')\, \left(\frac{E'-\Phi(r)}{E\,-\Phi(r)}\right)^n .
\end{align}
\end{subequations}

$\mu$ is the orbit-averaged diffusion coefficient in angular momentum, computed from the local diffusion coefficient in velocity (\ref{eq:DeltaV2per}) via double integration; unlike similar coefficients $A$ and $D$, here the averaging cannot be reduced to a single integral (except for the first term $I_0$)%
\footnote{Most early studies neglected the second term in this integral, corresponding to scattering by stars with higher binding energies; this underestimates the diffusion coefficient at low $|E|$, but the error is rather minor in the region of peak flux ($\lesssim 20\%$), and moreover, low $|E|$ corresponds to the full-loss-cone regime where the value of $\mu$ does not matter anyway.}.
These expressions rely on the standard relaxation theory, neglecting the effect of resonant relaxation in a nearly-Keplerian potential \citep{RauchTremaine1996}. However, as shown by recent detailed calculations, the overall impact of resonant relaxation on the loss rates is surprisingly moderate, because the enhancement of relaxation rate at intermediate angular momenta is compensated by a suppression of relaxation at very low angular momenta due to rapid relativistic precession \citep{Merritt2015b,Alexander2017}. Therefore, we retain the classical expressions for the diffusion coefficient in angular momentum.

The steady-state solution to the diffusion equation in angular momentum has a nearly logarithmic profile, determined by the boundary condition \citep{LightmanShapiro1977}. If $q \ll 1$ (the empty-loss-cone regime), $f\approx 0$ at the loss-cone boundary $\Rlc$, but since $\Rlc \ll 1$, $f(E,L)$ does not vary substantially over most part of the angular-momentum range, and in the opposite, full-loss-cone regime, $f$ is even closer to a constant ($L$-averaged) value. The timescale for establishing the steady-state profile is also much shorter than the timescale for the diffusion in energy, unless the initial distribution was strongly non-uniform over the entire range of $L$. Therefore, we may approximate the effect of a loss cone in the two-dimensional Fokker--Planck equation for $f(E,L)$ by an energy-dependent loss term $-\nu(E)\,f(E)$ in the one-dimensional equation (\ref{eq:FokkerPlanckDt}) for $f(E)$, responsible for the steady-state flux in the angular-momentum direction \citep[e.g.,][]{Merritt2013}%
\footnote{The quantity denoted by $\alpha$ here corresponds to $q/\xi$ in that paper, but the approximation for $\xi(q)$ quoted there actually refers to $\alpha$, not $\xi$ (clearly $\xi<1$ from its definition).}. The loss rate is given by
\begin{align}  \label{eq:loss_rate}
\nu &= \frac{\mu}{\alpha + \ln(1/\mathcal R_\mathrm{LC})} ,
\qquad  \alpha\approx (q^2+q^4)^{1/4} .
\end{align}

In the absense of the loss-cone effects, the evolution of the system is invariant w.r.t.\ simultaneous rescaling of time and stellar mass; however, the difference in boundary conditions between empty- and full-loss-cone regimes breaks this invariance.

The evolution of the entire stellar system is described by a coupled set of Fokker--Planck%
\footnote{As discussed by \citet{Chavanis2013}, the name ``Fokker--Planck equation'', strictly speaking, refers to a linear parabolic PDE with diffusion coefficients arising from an external thermal bath, whereas our Equation (\ref{eq:FokkerPlanck}) is a non-linear equation with the evolving DF itself entering the expressions for the diffusion coefficients. However, it is traditionally known by this name in the stellar-dynamical context.}
and Poisson equations, with the diffusion coefficients computed self-consistently from the DF itself. As in previous studies \citep[e.g.][]{Cohn1980}, we solve them in turn, first advancing the evolution of the DF $f_c$ of all components (the Fokker--Planck step), and then computing the overall density profile and the potential (the Poisson step). The advantage of using $h$ as the independent variable for the DF is important in the Poisson step, where the small adjustment of the potential preserves the DF (adiabatic invariance). Previous studies also used this scheme, but the DF was first expressed in terms of the action variables and then converted back to energy, which is completely unnecessary. Of course, we still need to construct the mapping between $\Phi$ and $h$ in the updated potential, since it enters indirectly the expressions for the diffusion coefficients (\ref{eq:I_0}-\ref{eq:K_h},\ref{eq:FokkerPlanckCoefs}) through $g(h)$. 
In some contexts, it may be useful to evolve the system in a fixed external potential (e.g., the Keplerian potential of the central black hole); in this case the recomputation of $\rho(r), \Phi(r)$ and $h(\Phi)$ is omitted, but the diffusion coefficients still need to be updated in the course of evolution, which is mandatory for the energy conservation.
More details about the numerical implementation are given in the Appendix \ref{sec:Appendix}.

\section{Conservation laws}  \label{sec:Conservation}

The flux-conservative formulation of the Fokker--Planck equation and its discretized version preserve the mass exactly (up to roundoff errors). Since we express $f$ as a function of $h$ and keep it fixed when solving the Poisson equation, it also conserves the mass exactly.

The energy conservation is a more subtle property. In a fixed potential and without source or sink terms, the Fokker--Planck equation alone implies the following evolution equation for the energy density $E(h) f(h,t)$, where in the multi-component case $f \equiv \sum_c f_c$:
\begin{subequations}
\begin{align}
\frac{\d [E(h) f(h,t)]}{\d t} &= -E(h) \sum_c \frac{\d \mathcal F_c(h,t)}{\d h} =
-\frac{ \d \mathcal F_E(h,t)}{\d h} ,\!\!  \label{eq:EnergyFluxDeriv}  \\
-\mathcal F_E &\equiv \sum_c \bigg[ -E \mathcal F_c - \frac{D}g\, f_c + A_c\,I_{0,c} \bigg],  \label{eq:EnergyFlux}
\end{align}
\end{subequations}
and the mass flux $\mathcal F_c$ of each component is defined by (\ref{eq:MassFlux}).
This rather weird-looking conservation law results from the fact that the diffusion coefficients contain integrals over the DF itself. Taking the derivative of $\mathcal F_E$ by $h$, the first term yields $E(h)\;\d \mathcal F(h)/\d h \,+\, \mathcal F(h)/g(h)$, and the remaining terms neutralize the second half of this expression, leaving only the product of $E$ and $\D f/\D t$. The first term in the energy flux corresponds to advective transport (energy carried by particles moving through phase space), and the remainder is the conductive flux representing non-local energy exchange through collisional relaxation.

In the case of evolving potential, it seems impossible to derive a local conservation law, but one may demonstrate the conservation of the total energy, generalizing the derivation presented in the appendix of \citet{Cohn1979} to the multi-component case with an external potential.
Define $\rho(r)$ to be the density generated by the combined DF of all stellar components $f$ via (\ref{eq:rho_from_f}), and let $\Phi_\star(r)$ be the potential corresponding to this density via the Poisson equation (\ref{eq:Poisson}). We will specialize to the case when an external potential represents the central black hole of mass $M_\bullet$, but the derivation is similar for an arbitrary external distributed mass profile. Hence, the total potential is given by $\Phi(r) = \Phi_\star(r) - G\,M_\bullet/r$.

Define the total kinetic $T$ and potential energy $W$ as
\begin{align}
T &\equiv \frac{3}{2} \int_{\Phi(0)}^0\!\! \D E\,f(E)\,h(E) \;=\; \frac{3}{2} \sum_c K_{h,c}(h=\infty) ,  \label{eq:KineticEnergy} \\
W &\equiv \frac{1}{2} \int_0^\infty \D r\; 4\pi r^2\; \rho(r)\;\Phi_\star(r)\;
+\; M_\bullet\,\Phi_\star(0).  \label{eq:PotentialEnergy}
\end{align}
For the potential energy to be finite, the density must be shallower (steeper) than $r^{-5/2}$ at small (large) radii, although in fact it must drop faster than $r^{-3}$ at large radii for the total mass to be finite. In the presence of the central black hole, this condition is stricter -- the stellar potential must be finite at origin (i.e., the density must be shallower than $r^{-2}$ -- same requirement as for $K_E$ to be finite).

When the stellar density $\rho(r,t)$ evolves with time, so does its associated gravitational potential $\Phi_\star(r,t)$; hence the rate of change of the potential energy can be written as
\begin{align}
\frac{\D W}{\D t} &= \frac{\D}{\D t}\bigg[
\frac{1}{2} \int_0^\infty \D r\; 4\pi r^2\; \rho(r,t)\; \Phi_\star(r,t) \;+\; M_\bullet\,\Phi_\star(0,t) \bigg]  \nonumber \\
&= \frac{1}{2} \int_0^\infty \D r\; 4\pi r^2\;
\bigg\{ \rho(r,t)\; \frac{\d \Phi_\star(r,t)}{\d t}\;\; +  \nonumber \\
&\quad +\;\frac{\d \rho(r,t)}{\d t}\; \Phi_\star(r,t) \bigg\}\;
 +\; \frac{\D\;\big[ M_\bullet\,\Phi_\star(0,t) \big]}{\D t}  \nonumber \\
&= \int_0^\infty \D r\; 4\pi r^2\; \rho(r,t)\; \frac{\d \Phi_\star(r,t)}{\d t} \;+\;
\frac{\D\;\big[ M_\bullet\,\Phi_\star(0,t) \big]}{\D t},   \label{eq:dWdt}
\end{align}
where we have integrated the second term in curly braces by parts twice, using the Poisson equation, demonstrating that the two terms have equal contribution to $\D W/\D t$.

The sum of energies of all stars is given by
\begin{subequations}  \label{eq:SumEnergy}
\begin{align}
\scE &\equiv \sum_c K_{E,c}(h=\infty)\; = \int_{\Phi(0)}^0 \D E\;f(E)\;g(E)\;E \\
&= \int_{\Phi(0)}^0 \D E \;f(E)\;E\;
   4\pi \int_0^{r_\mathrm{max}(E)}\!\!\!\!\!\!\!\! \D r\; 4\pi r^2\; \sqrt{2\big[ E - \Phi(r) \big]}  \nonumber\\
&= \int_0^\infty \D r\; 4\pi r^2\;
   4\pi\int_{\Phi(r)}^0 \D E\;f(E)\; \sqrt{2\big[ E - \Phi(r) \big]}\; E .  \nonumber
\end{align}
Now if we substitute $E = \Phi_\star(r) - \frac{GM_\bullet}{r} + \frac12 v^2$ in the last line, and recall the definition of $\rho$ (\ref{eq:rho_from_f}), then the integral splits into three parts. The first one is twice the potential energy of self-interaction between stars, i.e., the first term in (\ref{eq:PotentialEnergy}) that involves only the product of $\rho(r)\,\Phi_\star(r)$.
The last one is the kinetic energy of all stars $T$ (\ref{eq:KineticEnergy}). The middle term can be integrated by parts twice to yield $M_\bullet\,\Phi_\star(0)$. Hence
\begin{align}  \label{eq:E}
\scE = T + 2 W - M_\bullet\,\Phi_\star(0) .
\end{align}
\end{subequations}

On the other hand, the total energy of the system is given by
\begin{align}  \label{eq:U}
U \equiv T + W = \scE - W + M_\bullet\,\Phi_\star(0) .
\end{align}

In a virial equilibrium, $T = -W/2 = -U$, and hence $\scE = 3U - M_\bullet\,\Phi_\star(0)$. We now consider the case when the stellar potential evolves together with the DF, so that the correspondence between $h$ and $E$ depends on time, and there are possibly source or sink terms in the r.h.s. of equation (\ref{eq:FokkerPlanckDt}). First,
\begin{align*}
\frac{\D \scE}{\D t} &= \frac{\D}{\D t} \int_0^\infty \D h\; f(h,t)\; E(h,t)  \nonumber\\
&= \int_0^\infty\! \D h\; \frac{\d f(h,t)}{\d t}\;E(h,t) +
\int_0^\infty\! \D h\; f(h,t)\;\frac{\d E(h,t)}{\d t} . \nonumber
\end{align*}
The first term is converted into the energy flux through the boundary, using (\ref{eq:EnergyFluxDeriv}), plus the integral term describing the energy change associated with source or sink terms:
\begin{align}
\mathcal S_E \equiv\int_0^\infty \D h\;E(h,t)\; \sum_c\Big[s_c(h,t) - \nu_c(h,t)\,f_c(h,t)\Big] .
\end{align}
In the second term we expand $\rho(r)$ according to (\ref{eq:rho_from_f}), exchange the order of integration in $r$ and $h$, and replace $\d E/\d t$ with the time derivative of the total potential. \begin{align*}
\frac{\D \scE}{\D t} &=
-\mathcal F_E(h,t)\bigg|_{h=0}^{\infty} +\; \mathcal S_E \nonumber\\
&\quad+ \int_0^\infty \D r\; 4\pi r^2\; \rho(r,t)\; 
\left[\frac{\d\Phi_\star(r,t)}{\d t} - \frac{G}{r}\frac{\D M_\bullet}{\D t} \right]
\end{align*}
Noting that $\Phi_\star(0)=-\int_0^\infty \D r\,4\pi G r \rho(r)$ and invoking equation (\ref{eq:dWdt}), we obtain
\begin{align}  \label{eq:dEdt}
\frac{\D \scE}{\D t} = -\mathcal F_E(h,t)\bigg|_{h=0}^{\infty} + \mathcal S_E +
\frac{\D W}{\D t} - M_\bullet\frac{\D\Phi_\star(0,t)}{\D t}.
\end{align}
Finally, taking the time derivative of $U$ (\ref{eq:U}), we obtain
\begin{align}  \label{eq:dUdt}
\frac{\D U}{\D t} &=
\frac{\D\scE}{\D t} - \frac{\D W}{\D t} + \frac{\D\; \big[M_\bullet\,\Phi_\star(0,t) \big]}{\D t}
\nonumber\\
&= -\mathcal F_E(h,t)\bigg|_{h=0}^{\infty} + \mathcal S_E +
\frac{\D M_\bullet}{\D t}\,\Phi_\star(0).
\end{align}

In most cases, the energy flux $\mathcal F_E$ tends to zero as $h\to 0$ or $h\to\infty$, hence the total energy of the system is conserved in the absense of source/sink terms or changes in $M_\bullet$; however, we will later see that the situation is different in the case of a Bahcall--Wolf cusp.

\section{The Bahcall--Wolf cusp revisited}  \label{sec:BWcusp}

We now reconsider the classical problem of a steady-state stellar distribution around a central black hole of mass $M_\bullet$, focusing on the region inside the black hole sphere of influence (defined as the radius $r_\mathrm{infl}$ enclosing the mass of stars equal to $2M_\bullet$). Hence the potential is determined only by the black hole ($\Phi(r) = -GM_\bullet/r$), and the correspondence between energy and phase volume is given by (\ref{eq:Eofh_Kepler}).

Let us first neglect the loss of stars into the black hole, and consider a one-component system described by $f(h)$. In this case, as demonstrated by \citet{BahcallWolf1976}, the physically relevant solution has zero flux of mass $\mathcal F$ (\ref{eq:MassFlux}), which corresponds to $f(E) \propto (-E)^{1/4}$ or $f(h) \propto h^{-1/6}$. However, this solution cannot extend all the way to large radii, because the total mass is infinite; hence, we need to consider the effect of small deviations from a pure power law. Let
\begin{align}  \label{eq:f_perturb}
f(h) = C_0\,h^{-1/6} + C_1\,h^\mu
\end{align}
be the perturbed Bahcall--Wolf solution at small $h$; we mandate that $\mu>-1/6$ since the first term should dominate as $h\to 0$. Then, working out the functions $I_0, K_g, K_h$ and eventually the flux $\mathcal F$, we obtain that the dominant term is $\mathcal F(h) \propto C_1\,h^{\mu+1}\,f(h)$, hence from (\ref{eq:FokkerPlanckDt}) it follows that the time derivative of $f$ is $\propto C_1\,h^\mu\,f(h)$.
If the solution needs to stay self-similar, then $\d f/\d t \propto f$, and hence $\mu=0$. In this case the mass and energy fluxes are, to the leading order,
\begin{subequations}  \label{eq:flux_perturb}
\begin{align}
\mathcal F(h)   &\approx -\textstyle \frac{29}{20}\, \Gamma\,m\,C_1\, h\,f(h) \;\;
\propto\;\; -C_1\,h^{5/6}, \\
\mathcal F_E(h) &\approx \textstyle \frac{64}{25}\;3^{1/3}\,\pi^2\,G^2M_\bullet^2\;\Gamma\,m\,C_0^2 
+ 5\, E(h)\,\mathcal F(h)  \label{eq:EnergyFluxBW} \\
&\propto \mathrm{const} + C_1 h^{1/6},  \nonumber
\end{align}
and the corresponding density profile is
\begin{align}
\rho(r) &= \frac{2^{15/4}\,3^{1/6}\,\pi^2}{21\;\Gamma(\frac34)^2} \frac{(G\,M_\bullet)^{5/4}}{r^{7/4}}\,C_0 +
\frac{2^{7/2}\,\pi}{3} \frac{(G\,M_\bullet)^{3/2}}{r^{3/2}}\,C_1 .
\end{align}
\end{subequations}

As expected, the mass flux is proportional to $C_1$, i.e., vanishes in the case of a pure Bahcall--Wolf cusp. However, in the perturbed DF the mass flux is small but finite, and is directed inward (if $C_1>0$) or outward (if $C_1<0$); correspondingly, the DF increases or decreases with time while maintaining approximately the same functional form $\propto h^{-1/6}$. The energy flux, on the other hand, has a constant term that depends on the amplitude of the unperturbed DF ($C_0$) and corresponds to the energy being pumped into the system through the boundary at $h=0$. In other words, the black hole acts as a heat source, but the rate of energy production is determined by the stellar distribution itself. This was noticed already in the early papers (e.g., \citealt{BahcallWolf1976}, \S IIIe; \citealt{LightmanShapiro1977}, \S IIa), but attributed to the (negative) energy being carried into the black hole by captured stars (and therefore extracted from the system). As we see from (\ref{eq:EnergyFluxBW}), the energy flux remains finite even without any absorption of stars, i.e., even if $C_1=0$. It may be also understood as the heat conduction flux: the existence of the black hole mandates that the velocity dispersion (temperature) rises towards small radii as $r^{-1/2}$, and hence the energy is transported outwards. In this context, the black hole itself acts as a heat bath, because its own binding energy is formally infinite. The advective component of the energy flux ($E\mathcal{F} \propto h^{1/6}$) is sub-dominant to the conductive flux at high $|E|$ -- this is an essential feature of the Bahcall--Wolf solution (the dominant term in the mass flux that would be proportional to $h^{2/3}$ vanishes identically, leaving only the next-order correction $\propto h^{5/6}$). 

\begin{figure}
\includegraphics{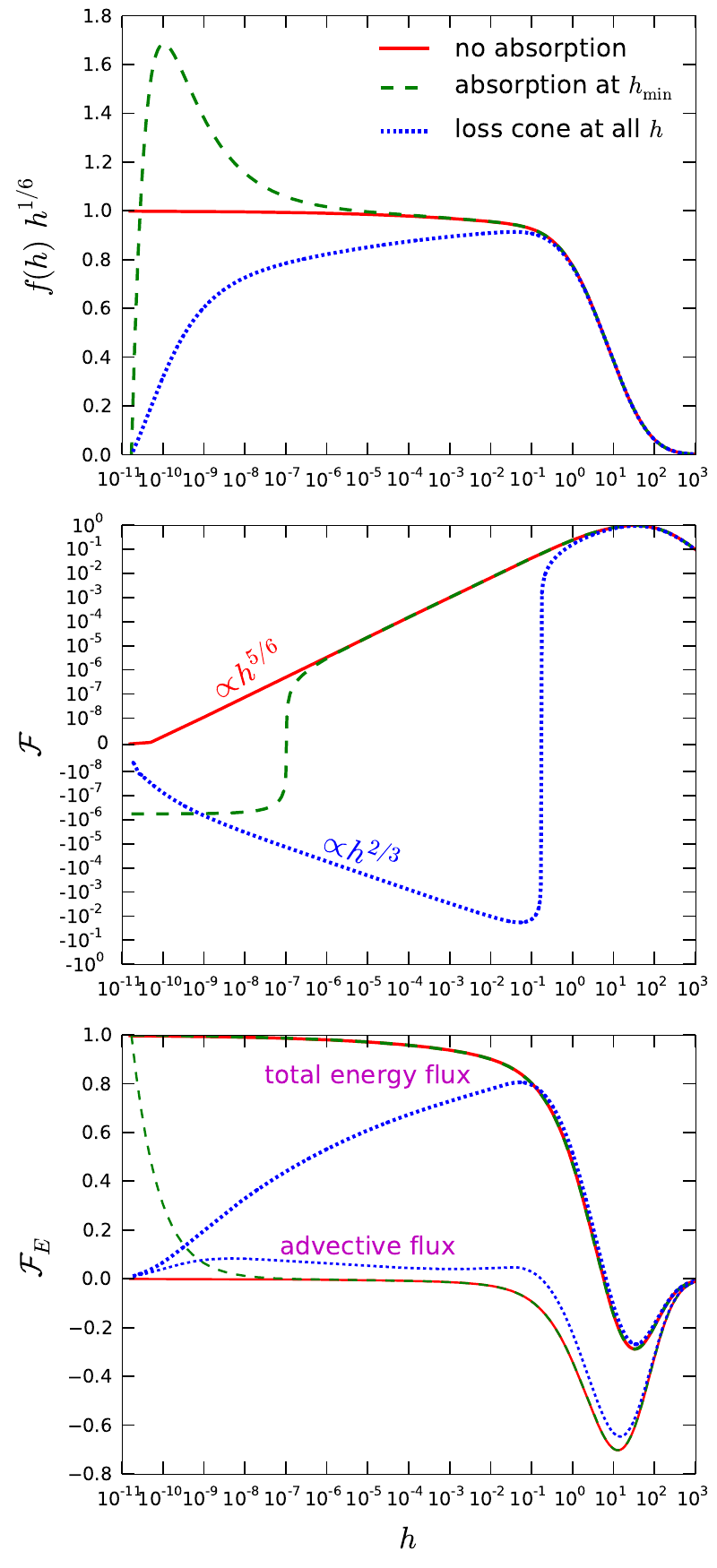}
\caption{The Bahcall--Wolf cusp in three variants: without captures (red solid curves), with an absorbing boundary at $\hmin\approx 10^{-11}$ (green dashed curves), and with loss-cone captures at all energies (blue dotted curves). $h=1$ corresponds to the energy $E=\Phi(r_\mathrm{infl})$, where $r_\mathrm{infl}$ is the influence radius that encloses the mass of stars equal to twice the black hole mass $M_\bullet$. \protect\\
Top panel: normalized DF $f(h)\,h^{1/6}$; middle panel: mass flux $\mathcal F(h)$ (\ref{eq:MassFlux}); bottom panel: energy flux $\mathcal F_E(h)$ (\ref{eq:EnergyFlux}), and separately the advective part of the energy flux $E\mathcal{F}$ in thinner lines. \protect\\
The initial model had a Plummer DF and a black hole with mass $M_\bullet=0.1$ times the stellar mass.
}  \label{fig:bwcusp}
\end{figure}

\begin{figure}
\includegraphics{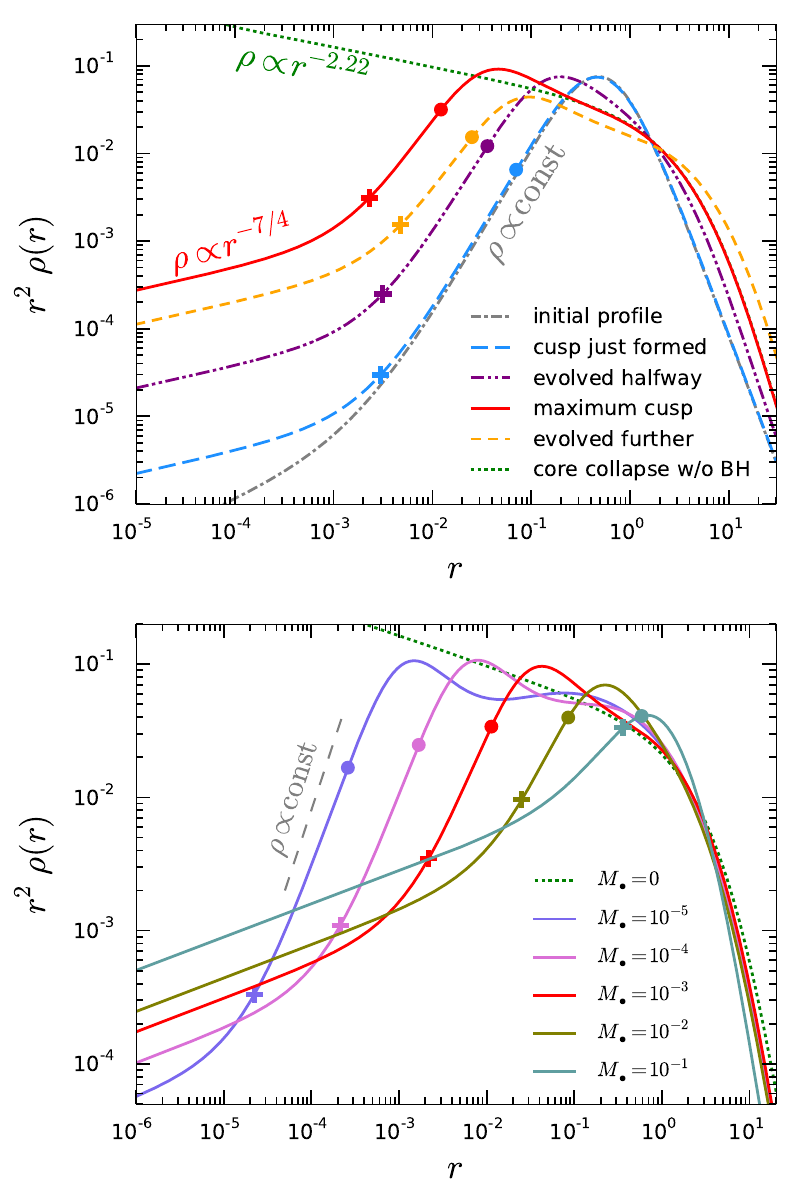}
\caption{Top panel: evolution of density profile in a Bahcall--Wolf cusp.\protect\\
Shown is the density multiplied by $r^2$ as a function of radius. Grey dot-dashed line is the original profile -- a Plummer sphere in virial units, with a central black hole of mass $M_\bullet=10^{-3}$ of the total mass; its presense adiabatically modifies the density profile at $r\lesssim 10^{-3}$ to form a weaker cusp $\rho\propto r^{-3/2}$.
Blue dashed line is the profile just after the Bahcall--Wolf cusp has formed, at $t\simeq 0.25\,T_\mathrm{r,h}$, where $T_\mathrm{r,h}$ is the half-mass relaxation time. Purple dot-dot-dashed line corresponds to $t=7.5\,T_\mathrm{r,h}$ and red solid line -- to $t=15\,T_\mathrm{r,h}$, at which point the cusp amplitude reaches its maximum. For comparison, a model without a black hole reaches a core collapse at this time, and its density profile is shown in dotted green line. Orange dashed line is the profile at $t=30\,T_\mathrm{r,h}$, where the cusp has decreased in amplitude as the energy is being pumped into the system, pushing the mass outwards.
Bullets mark the influence radius (containing the mass of stars equal to $2M_\bullet$), and crosses -- the cusp radius (defined as $r_\mathrm{cusp}\equiv GM_\bullet/\sigma^2$, where $\sigma$ is the velocity dispersion at this radius) at corresponding times.\protect\\
Bottom panel: density profiles of models with different black hole masses in the self-similar (post-core-collapse) regime. From left to right, models have $M_\bullet=10^{-5}$ to $10^{-1}$; for comparison, the core-collapsed model without a black hole (same as in the top panel) is shown by a dotted curve. Radii are normalized to the virial radius, i.e., all models have the same total energy. Bullets and crosses mark the influence and cusp radii.
}  \label{fig:bwcusp_evol}
\end{figure}

Of course, in reality the black hole imposes an absorbing boundary at a small but finite $\hmin$, so that $f(\hmin)=0$, and the flux of mass through the boundary is non-zero. Figure~\ref{fig:bwcusp} demonstrates that the DF is close to the solution without an absorbing boundary over a large range of $h$, but has a distinct hump at $\hmin < h \lesssim 10^3\hmin$ (top panel, green curve). The mass flux $\mathcal F$ is negative in this range, corresponding to the flow towards the black hole; however, outside this region the flux is still directed outwards, as in the case without absorption (middle panel). Moreover, the value of the energy flux $\mathcal F_E$ is virtually identical in these two situations (bottom panel), although the importance of different terms in (\ref{eq:EnergyFlux}) varies between these cases. In the case without an absorbing boundary, $E(h)\mathcal F(h) \propto h^{1/6}$ whereas the other two terms tend to a constant limit as $h\to 0$: the energy is transported by conduction, instead of being carried by the mass flow (advection). In the case of an absorbing boundary, the last two terms vanish at $\hmin$, since $f(\hmin)=0$ and $A\propto M(h<\hmin)=0$, and the remaining term $\mathcal F_E=E(\hmin)\,\mathcal F(\hmin)$ corresponds to the energy of captured stars removed from the system. However, in both cases $\mathcal F_E$ is ultimately determined by the maximum rate at which the energy can be transported into the outer parts of the system. The asymptotic expressions for the DF, fluxes and density (\ref{eq:f_perturb},\ref{eq:flux_perturb}) match the numerical solution very well in the range of $h \lesssim 0.1\hinfl$, where $\hinfl$ corresponds to the potential at the radius of influence.

Finally, if we account for the loss-cone effects by adding a loss term (\ref{eq:loss_rate}), the DF becames somewhat suppressed inside $\sim 0.1\hinfl$, and the mass flux is directed inwards in this range (however, it is still two orders of magnitude smaller than the maximum outward flux at roughly the half-mass radius). The rate of energy extraction from the system by captured stars is spread across the entire cusp range, hence the outward energy flux gradually decreases towards $\hmin$, but the total energy change rate $\D U/\D t$ is roughly the same as without the loss-cone effects. In all cases, the advective component of the energy flux is sub-dominant, as illustrated in the bottom panel of Fig.~\ref{fig:bwcusp}.

It is important to note that the relative change of the total energy of the stellar system $U$ is far larger than the change of the total mass (the latter is zero without an absorbing boundary, practically negligible in the case of absorption at $\hmin$, and still fairly small if loss-cone effects are included). Hence, the system gradually expands, responding to the energy source at its center. Following \citet{Henon1975}, the long-term asymptotic behaviour of an isolated stellar system without mass loss may be described by a self-similar (homologous) model. If we denote the characteristic radius by $\tilde r$, the total energy $U\propto GM/\tilde r$, and the relaxation time $T_\mathrm{rel} \propto \tilde r^{3/2} \propto |U|^{-3/2}$. Since the heat production rate in the cusp is determined by the maximum energy flux that can be transported through the system, $\mathcal F_E \propto U / T_\mathrm{rel}$, the total energy decreases with time as $\D U/\D t = \mathcal F_E \propto |U|^{5/2}$, and hence $U(t) \propto t^{-2/3}$.

Interestingly, the amplitude of the cusp may either increase or decrease with time, depending on the sign of $C_1$ in (\ref{eq:f_perturb}). Figure~\ref{fig:bwcusp_evol}, top panel, shows the evolution of density profiles of a model with a Plummer DF and a central black hole of mass $M_\bullet=10^{-3}\,M$. At first, the Bahcall--Wolf cusp develops at small radii and gradually fades into the cored density profile. As the core radius shrinks and its density increases, so does the amplitude of the cusp, maintaining roughly the $r^{-7/4}$ profile inside a fraction of influence radius. At the time when an equivalent isolated system would reach a core collapse, the density in the cusp is maximal, but much less steeply rising than in a core-collapsed model \citep[e.g.][]{Baumgardt2005}. The subsequent evolution follows a self-similar profile with a gradually decreasing density and proportionally increasing radius. The ratio between the cusp radius and the core radius is $\propto \smash{M_\bullet^{1/4}}$ for sufficiently small $M_\bullet$, as shown by \citet{Heggie2007} from simple dimensional arguments (Figure~\ref{fig:bwcusp_evol}, bottom panel); however, this only holds in the post-collapse phase (i.e., this ratio is clearly not constant in the top panel).
The fact that the Bahcall--Wolf cusp evolves in amplitude has not been observed in the early papers, which produced a steady-state solution by fixing the value of the DF at the outer cusp boundary. In reality, this external heat bath must be replaced by a real physical system which responds to the heat source at the center.

\section{Validation of the Fokker--Planck approach}  \label{sec:Validation}

\begin{figure}
\includegraphics{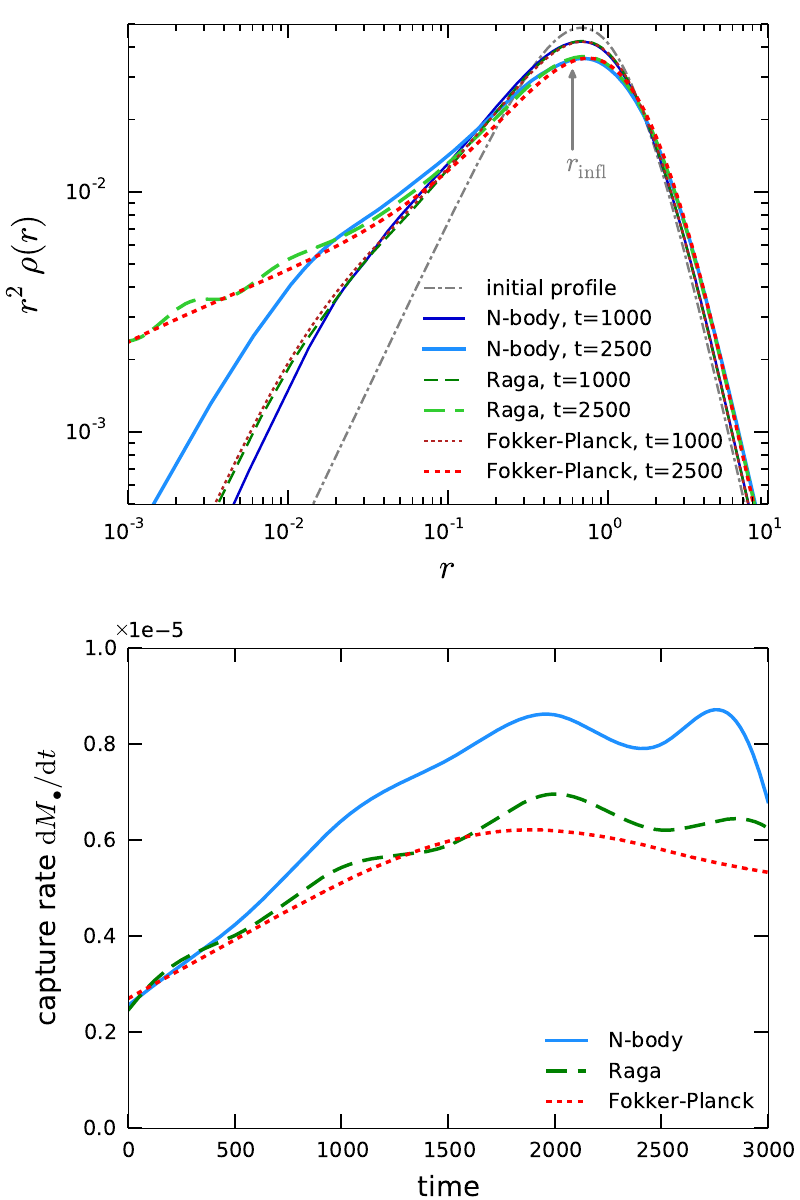}
\caption{Comparison of three methods for studying the evolution of a stellar cusp around a massive black hole: direct $N$-body simulation (solid blue curves), the Monte Carlo code \textsc{Raga} (dashed green), and the Fokker--Planck approach from this paper (dotted red).
The initial system had a density profile (\ref{eq:ZhaoProfile}) with the inner slope $\gamma=0.6$, total mass $M=1$ and a black hole of mass $M_\bullet=0.1$, with a capture radius $r_\mathrm{LC}=10^{-5}$. \protect\\
Top panel shows the density at three different moments of time: initial (grey), $t=1000$ (thinner and darker curves), and $t=2500$, when the Bahcall--Wolf cusp is fully in place (thicker and lighter curves). The agreement between Monte Carlo and Fokker--Planck models is very good, while in the $N$-body model the steep cusp only extends down to $r\simeq 0.02\simeq 0.03\,r_\mathrm{infl}$, and becomes shallower inwards. This could be attributed to the resonant relaxation which is not accounted for in the other two methods.\protect\\
Bottom panel shows the evolution of capture rate as a function of time: it increases roughly twofold when the cusp is in place, and then slowly drops as the density continues to decrease in amplitude. Again the $N$-body model has a somewhat higher rate owing to the resonant relaxation.
}  \label{fig:bwtest}
\end{figure}

To demonstrate how well does the Fokker--Planck description match the actual evolution of a stellar system, we compare it to two other methods: the stellar-dynamical Monte Carlo code \textsc{Raga} \citep{Vasiliev2015} and the $N$-body code $\phi$\textsc{grape}ch \citep{Harfst2008}. Both methods represent the system as a collection of discrete particles, as opposed to the description in terms of smooth functions $f(h)$ and $\Phi(r)$ in the Fokker--Planck approach. However, the evolution is treated quite differently in these codes. $\phi$\textsc{grape}ch is a conventional direct-summation code with GPU acceleration provided by the \textsc{Sapporo} library \citep{Gaburov2009} and chain regularization for an accurate treatment of particle encounters with the central black hole. The potential computed from particles represents both the smooth global profile and the fluctuations driving the collisional relaxation. By contrast, in the Monte Carlo method the global potential is represented as a smooth function of radius, which is computed from particle positions, but has a much lower noise due to several spatial and temporal smoothing techniques. The effect of two-body relaxation is simulated by adding perturbations to particle velocities as they move in the smooth potential; the amplitude of these perturbations follow the same prescription as in the Fokker--Planck approach, but without orbit-averaging. The actual DF represented by particles does not need to be isotropic in the Monte Carlo method; however, in computing the diffusion coefficients, an isotropic approximation is employed.

\citet{Vasiliev2015} demonstrated that the growth of the Bahcall--Wolf cusp is well described by the Monte Carlo approach, in comparison to the direct $N$-body simulation. We now augment this comparison to include the loss-cone effects.
Particles approaching to within a given distance $r_\mathrm{LC}$ from the black hole are captured, and a certain fraction of their mass is added to the black hole mass; in this test we adopt the accretion fraction of 100\%, even though in reality it is likely much smaller than unity \citep{MetzgerStone2016}. Thus the capture boundary is given in physical space in both the Monte Carlo and the $N$-body approaches, whereas the Fokker--Planck formulation adopts a more approximate prescription in terms of angular-momentum boundary $L_\mathrm{LC} \equiv \sqrt{2 G M_\bullet\,r_\mathrm{LC}}$ and a steady-state expression for the loss-cone flux.

We take the initial density to be described by a general double-power-law profile \citep{Zhao1996}:
\begin{subequations}  \label{eq:ZhaoProfile}
\begin{align}
\rho(r) &\equiv \rho_0\; \left(\frac{r}{r_0}\right)^{-\gamma}
\left[1 + \left(\frac{r}{r_0}\right)^\alpha \right]^{(\gamma-\beta)/\alpha} , \\
\rho_0 &\equiv \frac{M}{4\pi\,r_0^3} \; \frac{\alpha\,\Gamma\big(\frac{\beta-\gamma}\alpha\big)}
{\Gamma\big(\frac{3-\gamma}\alpha\big)\; \Gamma\big(\frac{\beta-3}\alpha\big) } .
\end{align}
\end{subequations}

In the Kepler potential, an isotropic DF cannot have a density profile shallower than $r^{-1/2}$. For this test, we adopt $\gamma=0.6$, $\beta=5$, $\alpha=2$, the black hole mass $M_\bullet = 0.1\,M$, and the capture radius $r_\mathrm{LC}=10^{-5}\,r_0$. The $N$-body simulation has $N=65535$ equal-mass particles plus the black hole, and to compare the evolution rate with the other methods, we set the Coulomb logarithm to $\ln\Lambda \simeq \ln(M_\bullet/m_\star) \approx 9$. The values of $N$ and $r_\mathrm{LC}$, of course, are far from realistic for galactic nuclei.
If both the the relaxation time and the total simulation time are multiplied by $K$, and the capture radius -- by $K^{-1}$, this preserves the loss-cone filling factor $q$ (\ref{eq:q_losscone}), hence the captured mass per relaxation time remains almost the same, up to a logarithmic correction in (\ref{eq:loss_rate}). In fact the adopted values roughly correspond to one of the models of the Milky Way nucleus from the next section, after rescaling by $K=350$ (i.e., taking $N=4\times10^7,\; \ln\Lambda=15,\; r_\mathrm{LC}=3\times 10^{-8}\,r_0$ and setting $r_0=5$~pc, which makes one time unit equivalent to $0.9\times10^7$~yr). 

Fig.~\ref{fig:bwtest} compares the density profiles at different times and the capture rates between three methods. It is clear that the Fokker--Planck and Monte Carlo approaches result in a very similar evolution, which is not surprising because both are based on the same prescription for relaxation. The approximate treatment of the loss cone in the Fokker--Planck method appears to be sufficiently accurate. By contrast, in the $N$-body system the cusp does not extend all the way to the center, although the density profiles match those of the other methods at radii $\gtrsim 0.03\,r_\mathrm{infl}$. At smaller radii, the enhancement of angular-momentum diffusion due to resonant relaxation leads to a more rapid loss of stars, preventing the growth of the cusp. However, the total capture rate increases only moderately ($\lesssim 50\%$), in line with the earlier studies \citep{RauchTremaine1996,HopmanAlexander2006}. Moreover, in our $N$-body simulations we neglected relativistic effects, which quench the resonant relaxation at high eccentricities and counteract its impact on the density profiles \citep{Merritt2015b}.

Overall, the agreement between the approaches is satisfactory, taking into account various approximations made in the Monte Carlo and Fokker--Planck methods. It should be noted that the $N$-body simulation took a few days, the Monte Carlo simulation -- a few CPU hours, and the Fokker--Planck run -- only a few minutes.

\section{The Milky Way nuclear star cluster}  \label{sec:MilkyWayNSC}

We now apply the Fokker--Planck method to construct evolutionary models of the nuclear star cluster (NSC) of our Galaxy. At present, it hosts a central black hole of mass $M_\bullet \approx 4\times10^6\,M_\bullet$ \citep{Boehle2016}, and the surface brightness profile of the old stellar population moderately rises towards the center \citep{Schoedel2017}. The slope of the density profile is somewhat lower than expected for a steady-state Bahcall--Wolf profile, but on the other hand, it does not seem to have a central depression (core), as inferred in earler studies \citep[e.g.,][]{Do2009}.

We consider one- and two-component models with initial density profile described by (\ref{eq:ZhaoProfile}). The total mass of the NSC is taken to be $2.5\times 10^7\,M_\odot$ \citep{Schoedel2014}, and we keep it fixed throughout the evolution, postponing the role of star formation for a later study (Generozov et al., in prep.). 
For the one-component models, we assume equal-mass stars with $m_\star=1\,M_\odot$, and for the two-component models, we take the stellar-mass black holes of $m_\mathrm{h}=10\,M_\odot$ to contribute 1\% to the total mass of NSC \citep{Alexander2005}, distributed initially with the same profile. The capture radius is set to $r_\mathrm{LC}=(M_\bullet/m_\star)^{1/3}r_\star \approx 3.6\times10^{-6}$~pc for solar-type stars, and $r_\mathrm{LC}=8GM_\bullet/c^2 \approx 1.5\times10^{-6}$~pc for black holes; to facilitate the comparison between models, we neglect the growth of $M_\bullet$ due to accreted mass.

\begin{figure}
\includegraphics{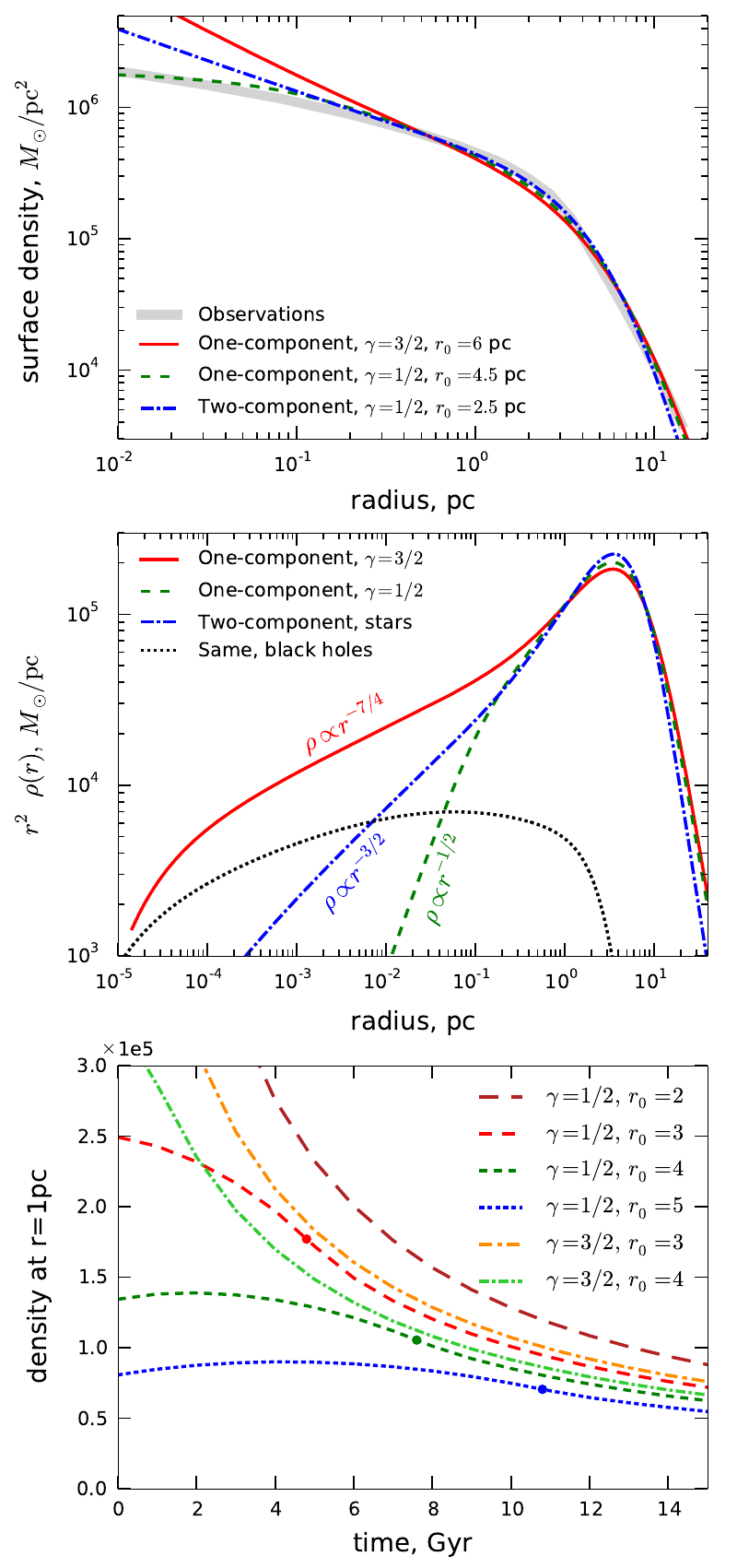}
\caption{Density profiles of the Milky Way nuclear star cluster.\protect\\
Top panel: surface density profile as a function of projected radius.
Grey curve shows the observational data from \citet{Schoedel2017}, normalized to have the 3d density $\rho=1.1\times10^5\,M_\odot/\mbox{pc}^3$ at $r=1$~pc; other lines show models that have the same density at 1~pc at time $t=10$~Gyr. Red solid curve is the one-component model with a steep initial density profile, which has formed a Bahcall--Wolf cusp in less than 1~Gyr. Green dashed line is a one-component model with an initially depleted profile, which did not have enough time to re-grow the cusp. Blue dot-dashed line is a similar two-component model, in which the heavy stellar-mass black holes accelerate the formation of the quasi-stationary profile. \protect\\
Middle panel: 3d density profile of the same models, multiplied by $r^2$ to compress the dynamic range. Additionally, the density profile of the heavy species in the two-component model is shown by black dotted curve. \protect\\
Bottom panel: evolution of the stellar density at $r=1$~pc for several two-component models with different initial cusp slopes and scale radii. Dots mark the approximate formation time of the cusp, after which the evolution proceeds self-similarly (i.e., different models have the almost the same profiles, but attained at different times).
}  \label{fig:mwnsc_dens}
\end{figure}

We explore two one-parameter families of models, with the inner cusp slope $\gamma=1/2$ or $3/2$, varied scale radius $r_0$, and other parameters fixed to $\alpha=2$, $\beta=5$. The former choice is the shallowest possible cusp slope for an isotropic model, and the initial DF is very strongly suppressed at all energies inside the sphere of influence; this kind of initial conditions mimics the depletion of the cusp resulting from a previously existing binary SBH \citep[e.g.,][]{MerrittSzell2006,Merritt2010}. The other choice corresponds to an adiabatically grown cusp around a SBH (it is very similar to the density profile obtained by embedding a SBH into a pre-existing Plummer model and adiabatically readjusting the potential while keeping $f(h)$ fixed, e.g., \citealt{Young1980,Quinlan1995}). The two sets of models are intended to bracket more realistic cases: in the former case, we expect that the re-growth of the cusp takes a significant time, while in the latter case the relaxation time is shortest and roughly constant inside $r_\mathrm{infl}$.

The evolution of all models follows a similar route: first a Bahcall--Wolf cusp grows from outside in, and subsequently the system reaches a self-similar expansion regime powered by energy transfer from the SBH. The density profiles at the latter stage have a nearly universal shape: as its amplitude gradually decreases, the characteristic scale proportionally increases. In single-component models with an initially shallow density profile ($\gamma=1/2$), the formation of the cusp takes longer than the Hubble time, unless the initial scale radius was unrealistically small. By contrast, in two-component models the evolution proceeds much faster due to rapid mass segregation, and they reach a self-similar regime much earlier, regardless of the initial density slope. The density of the heavy species (stellar-mass black holes) roughly follows a $\rho\propto r^{-7/4}$ profile at smallest radii, becoming steeper further out when they cease to dominate in the energy exchange rate (the so-called strong mass segregation regime, \citealt{AlexanderHopman2009, PretoAmaroSeoane2010}), while the lighter stars follow a somewhat shallower $\rho\propto r^{-3/2}$ profile inside $r_\mathrm{infl}$.

Figure~\ref{fig:mwnsc_dens} shows the density profiles of several models at the time $t=10$~Gyr, compared to the observations from \citet{Schoedel2017}. The classical Bahcall--Wolf cusp is clearly excluded by the data, as it produces much too steep projected density profiles at all radii inside $r_\mathrm{infl}$. Taken face value, a single-component model with an initially shallow profile and a suitably chosen initial scale radius matches the data best: the radius of this shallow ``core'' shrinks with time and reaches a value $\sim 0.1-0.2$~pc, just enough to produce a good fit to the surface density in the entire range $10^{-2}-10$~pc. This was the conclusion reached by \citet{Merritt2010}, who used a somewhat different initial profile, also with a shallow core.
However, single-component models are not particularly realistic. When we consider two-component models with only 1\% contribution of stellar black holes, the situation is very different: the cusp re-grows and a self-similar stage is achieved in a much shorter time, less than 10~Gyr, unless the initial scale radius was larger than $\sim 5$~pc (but in that case the present-day density is too low anyway). There is much less difference between models with initially shallow ($\gamma=1/2$) or steep ($\gamma=3/2$) cusps: the bottom panel of Figure~\ref{fig:mwnsc_dens} demonstrates that they all reach the same self-similar asymptotic regime, although at different moments of time (e.g., the model with initial $r_0=4$~pc and $\gamma=1/2$ has the same density profile at $t=8$~Gyr as the model with $r_0=3$~pc, $\gamma=1/2$ at $t=10$~Gyr, or the model with $r_0=3$~pc, $\gamma=3/2$ at $t=11$~Gyr, and their evolution is almost identical at all later times).
As the top panel shows, their projected density profile is still too steep at $r\lesssim 0.1$~pc compared to the observations. Our preliminary tests indicate that adding a moderate amount of continuous star formation at $r\gtrsim 1$~pc drives the resulting density profile very close to the observed one; we do not report these models here, but defer them to a separate paper.

Overall, the models with the present-day density at $r=1$~pc in the range $(0.8-1.5)\times10^5\,M_\odot/\mathrm{pc}^3$ agree rather well with the observed density profile (suitably scaled in amplitude), especially in the middle of this range. The enclosed mass within 1~pc is $(0.6-1)\times10^6\,M_\odot$, and the half-mass radius is $4-4.5$~pc, again in good agreement with observations \citep[e.g.][]{Schoedel2014}. As the middle panel of Figure~\ref{fig:mwnsc_dens} shows, the density of stellar-mass black holes is higher than that of stars inside $r\lesssim 10^{-2}$~pc, which has important implications for the rate of extreme-mass-ratio inspirals (EMRI) in Milky Way-sized galactic nuclei \citep{AmaroSeoanePreto2011}.
The present-day rates of black hole captures and stellar tidal disruptions are close to $5\times10^{-6}$ and $6\times10^{-5}$ events per year, respectively; these values are probably more sensitive to the model assumptions (spherical symmetry, neglect of resonant relaxation) than the inference about the global structure and density profile.

\section{Discussion and conclusions}  \label{sec:Summary}

We have reviewed the Fokker--Planck approach for studying the evolution of spherical isotropic stellar systems driven by collisional relaxation, and presented a publicly available%
\footnote{The Fokker--Planck solver is provided as part of the \textsc{Agama} library for galaxy modelling (Vasiliev, in prep.), available at \url{https://github.com/GalacticDynamics-Oxford/Agama}.}
code \textsc{PhaseFlow}, which can handle multi-component systems with star formation and loss-cone effects. A novel aspect in this work is the use of phase volume instead of energy as the argument of the distribution function, which facilitates the solution. We discussed the energy conservation and transport properties of the system and constructed a perturbative solution for the distribution of stars around a central massive black hole (the Bahcall--Wolf cusp). Despite being commonly labelled as a ``steady-state solution'', it actually evolves with time, following the energy transfer from the black hole to the stellar system. At late times, the system approaches a self-similar expansion regime powered by the central heat source \citep[e.g.,][]{MarchantShapiro1980,Merritt2009}.

We applied the method to the nuclear star cluster of the Milky Way and demonstrated that in the presence of two mass components (lighter stars and heavier stellar-mass black holes), mass segregation leads to accelerated formation of the cusp within a few gigayears, and the subsequent evolution occurs in a self-similar regime with little dependence on initial conditions. The present-day density in the model matches quite well the observed surface brightness profile in the range $0.1-10$~pc, although is somewhat steeper further in. We conjecture that the inclusion of star formation predominantly concentrated at radii $\gtrsim 1$~pc \citep[e.g.,][]{AharonPerets2015} would bring the present-day profile into better agreement with observations. The recent $N$-body model of \citet{Baumgardt2017}, which included star formation and realistic mass spectrum, looks quite similar to our simplified two-component Fokker--Planck models, which take only a few seconds to a few minutes to run, allowing a comprehensive exploration of parameter space.

Of course, the one-dimensional Fokker--Planck description is only valid for spherically-symmetric systems with isotropic velocity distribution. How serious is this limitation depends on the problem. 
For instance, angular-momentum relaxation in axisymmetric systems occurs faster, leading to a few-fold increase in the loss-cone flux, as demonstrated by \citet{VasilievMerritt2013} using a two-dimensional Fokker--Planck equation, neglecting the diffusion in energy and assuming a fixed potential of the black hole. Resonant relaxation also leads to a faster diffusion in angular momentum, but taking into account its suppression due to relativistic precession of high-eccentricity orbits, the overall impact on the evolution and the loss-cone capture rate is rather moderate \citep{Merritt2015b}. In computing the loss-cone flux, we assumed a steady-state profile of the DF in angular momentum, which is established after a small fraction of the relaxation time; at earlier times, the flux may be higher or lower, depending on the initial anisotropy profile \citep{WangMerritt2004, LezhninVasiliev2015, Stone2017}. The loss of stars into the black hole has very little impact on the global evolution of the system, which is driven by energy relaxation, therefore these details are immaterial unless we are interested specifically in the rate of tidal disruption events. The DF anisotropy may also be important in the context of tidal mass loss, necessitating a full 2d Fokker--Planck treatment \citep[e.g.][]{TakahashiBaumgardt2012}.

Fokker--Planck codes for axisymmetric systems are limited to the two-integral case \citep{Goodman1983,EinselSpurzem1999}, and none exist for triaxial systems. On the other hand, collisional relaxation in these systems can be studied using the more general Monte Carlo approach \citep{Vasiliev2015}, which is however much more computationally demanding, or still more expensive direct $N$-body simulations. We believe that the Fokker--Planck method still remains valuable and could be used as a quick tool to explore a large variety of models and determine general trends, complementing the more elaborate approaches. We hope that the software described in this paper and provided to the community will facilitate the applications of this method in various contexts.

This work was supported by the European Research council under the 7th Framework programme  (grant No.\ 321067) and by NASA (grant No.\ NNX13AG92G). I thank A.Generozov for valuable comments on the early draft and continuous feedback on the code.


\appendix

\section{Numerical method}  \label{sec:Appendix}
Here we present our formulation of finite-element method for the Fokker--Planck equation and show its relation to the classical Chang\&Cooper scheme (which is also included as a special case).

We start by defining a scaled spatial coordinate $x$ instead of $h$, such that the integrals involving the DF are written as $\int f(x)\, \mu(x)\,\D x$, where $\mu \equiv \D h(x)/\D x$. More generally, we define the inner product in the space of all functions of $x$ in the domain $x_-\le x\le x_+$ as
\begin{align}
\langle f(x), g(x) \rangle \equiv \int_{x_-}^{x_+} f(x)\,g(x)\;\mu(x)\,\D x.
\end{align}

We work with a finite-dimensional subspace of functions $\tilde f(x)$ that can be represented in a discretized form:
\begin{align}
\tilde f(x) \equiv \sum_{j=1}^B f_j\,\hat e_j(x),
\end{align}
where $\hat e_j(x)$ are fixed basis function and $f_j$ are expansion coefficients. To find the coefficients $f_j$ that best describe the discretized counterpart of an arbitrary continuous function $f(x)$, we demand that
\begin{align}
\mathcal P_i\{f\} \equiv \big\langle f(x), \hat e_i(x) \big\rangle\; =\;
\big\langle \tilde f(x), \hat e_i(x) \big\rangle =
\sum_{j=1}^B M_{ij}\,f_j \quad \mbox{for all }i=1..B,\;\;
\mbox{ where }M_{ij} \equiv \langle \hat e_i, \hat e_j\rangle ;
\end{align}
in other words, the projection of function $f(x)$ onto each basis vector is the same as the projection of its discrete counterpart.
This linear system may be written more compactly as $\mathsf{M}\, \boldsymbol{f} = \boldsymbol{\mathcal P}\{f\}$ (denoting vectors with boldface and matrices with sans-serif font).

The Fokker--Planck equation (\ref{eq:FokkerPlanck}) for a given species reads
\begin{align}
\frac{\d f(x,t)}{\d t} \;=\; -\frac{1}{\mu(x)} \frac{\d \mathcal F(x,t)}{\d x} + s(x,t) - \nu(x,t)\,f(x,t) \,, \qquad -\mathcal F(x,t) \equiv A\big(x\big)\,f(x,t) + \frac{D\big(x\big)}{\mu(x)}\,\frac{\d f(x,t)}{\d x},
\end{align}
where $A$ and $D$ are the advection and diffusion coefficients (\ref{eq:FokkerPlanckCoefs}), $s$ is the source term (star formation rate), $\nu$ is the loss-cone draining rate (\ref{eq:loss_rate}). We now apply the Galerkin projection operator $\mathcal P_i\{\circ\}\equiv \langle \circ,\hat e_i\rangle$ to both sides of this equation and replace $f(x,t)$ with its discretized representation $\tilde f(x,t) \equiv \sum_j f_j(t)\,\hat e_j(x)$:
\begin{align}
\sum_{j=1}^B M_{ij} \frac{\D f_j(t)}{\D t} =
-\int \hat e_i(x)\,\frac{\d\mathcal F(x,t)}{\d x}\,dx + s_i(t) - \sum_{j=1}^B V_{ij}(t)\,f_j(t) \,, \quad
s_i \equiv \mathcal P_i\{s\}, \quad V_{ij} \equiv \mathcal P_i\{\nu \hat e_j\}.
\end{align}
The first term may be integrated by parts to yield
\begin{align*}
\Big[ -\mathcal F(x,t)\,\hat e_i(x) \Big]\bigg|_{x_-}^{x_+} +
\int_{x_-}^{x_+} \mathcal{F}(x,t)\,\frac{\D \hat e_i(x)}{\D x} \,=\,
\Big[ -\mathcal F(x,t)\,\hat e_i(x) \Big]\bigg|_{x_-}^{x_+} + \:
\sum_{j=1}^B R_{ij}\,f_j \;,
\end{align*}
\begin{align}
R_{ij} \equiv -\int_{x_-}^{x_+} \left( A(x)\,\frac{\D\hat e_i(x)}{\D x}\,\hat e_j(x) +
\frac{D(x)}{\mu(x)}\,\frac{\D\hat e_i(x)}{\D x}\,\frac{\D\hat e_j(x)}{\D x} \right)\,\D x .
\end{align}
The expression in square brackets contains the flux through the boundaries of the integration region. For simplicity, we consider only two cases: (a) Neumann boundary condition with $\mathcal F=0$, in which case this term vanishes, or (b) Dirichlet boundary condition with $f(x_-,t)=0$. For our choice of basis functions (see below), $\hat e_i(x_-) = 1$ if $i=1$ and 0 otherwise; hence the coefficient $f_1$ is identically zero and may be excluded from the linear system, whereas for $i>1$ the boundary term vanishes again.

In the matrix form, the Fokker--Planck equation is a first-order differential equation for the vector of coefficients $\boldsymbol f$:
\begin{align}  \label{eq:FokkerPlanckMatrix}
\mathsf{M}\,\frac{\D \boldsymbol f(t)}{\D t} = \big( \mathsf{R} - \mathsf{V} \big) \boldsymbol{f}(t) + \boldsymbol{s}(t) .
\end{align}

The key point of the finite-element method is that the basis functions $\hat e_j(x)$ are nonzero only in a narrow range of $x$ each, so that their products entering the matrices $\mathsf{M,R,V}$ are nontrivial only if $|i-j| \le N$, where $N$ is of order a few. Then the matrices can be efficiently inverted using $LU$-decomposition with $\mathcal O(N^2B)$ operations (a familiar special case is the tridiagonal matrix algorithm for the case $N=1$). A suitable choice for the basis are B-spline functions -- piecewise polynomials of degree $N$, defined by a set of grid knots $x_k$, $k=1..K$; the total number of basis elements is $B=K+N-1$, each function is nonzero on at most $N+1$ consecutive intervals between knots, and has $N-1$ continuous derivatives at each knot.
In the case $N=1$, the basis element $\hat e_k$ is a $\wedge$-shaped function spanning two adjacent grid cells (from $x_{k-1}$ to $x_{k+1}$), and the expansion coefficients coincide with the values of the function at each knot ($\tilde f(x_k) = f_k$), but in general this does not hold.
To compute the integrals entering the projection operator $\boldsymbol{\mathcal P}$ and matrix elements, it is convenient to use the Gauss--Legendre quadrature with $N+1$ points per each segment $x_k\,..\,x_{k+1}$, which gives an exact result if the integrand is a polynomial of degree $\le 2N+1$ (in particular, a product of two basis functions).

The conventional finite-difference scheme may also be reformulated in the form of the matrix equation (\ref{eq:FokkerPlanckMatrix}) as follows. The basis functions are non-overlapping $\sqcap$-shaped blocks spanning intervals $x_{k-1/2}\,..\,x_{k+1/2}$ around each grid node $x_k$, where the half-indexed points are centers of grid cells $x_{k+1/2} \equiv (x_k+x_{k+1})/2$. The coefficients $f_k$ correspond to the nodal function values $\tilde f(x_k)$, the projection operator $\mathcal P_k\{f\} = (x_{k+1/2}-x_{k-1/2})\,f(x_k)$, and the matrices $\mathsf{M,V}$ are diagonal with elements $M_{ii} = (x_{i+1/2}-x_{i-1/2})\,\mu(x_i)$, $V_{ii}=M_{ii}\,\nu(x_i)$. The combined advection/diffusion matrix $\mathsf{R}$ is tridiagonal, and its elements are computed using the \citet{ChangCooper1970} prescription:
\begin{align*} 
R_{i\:i-1} &=  W^-_{i-1/2}\, C_{i-1/2},\quad
R_{i\:i}    = -W^-_{i+1/2}\, C_{i+1/2} - W^+_{i-1/2}\, C_{i-1/2},\quad
R_{i\:i+1}  =  W^+_{i+1/2}\, C_{i+1/2},  \\
C_{i+1/2}  &\equiv \frac{1}{(x_{i+1}-x_i)}\frac{D(x_{i+1/2})}{\mu(x_{i+1/2})}, \quad
w \equiv \frac{A(x_{i+1/2})}{C_{i+1/2}}, \quad
W^-_{i+1/2} \equiv \frac{w}{\exp w-1}, \quad W^+_{i+1/2} \equiv W^-_{i+1/2}+w .
\end{align*}
The idea behind these expressions is that in a near-equilibrium system, the advection and diffusion terms in the flux $\mathcal F(x_{i+1/2})$ nearly cancel each other; to make a finite-difference estimate more accurate, we take a suitably weighted combination of $f_i$ and $f_{i+1}$ instead of a simple-minded equal-weight average (see \citealt{ParkPetrosian1996} for mathematical details and a comparison of methods). In the finite-element method with stencil width $N>1$, the flux is estimated more accurately, and these intricacies are unnecessary.

It remains to devise a time integration strategy for equation (\ref{eq:FokkerPlanckMatrix}).
The time derivative in the left-hand side is replaced with the finite-difference approximation $\D\boldsymbol{f}/\D t = (\fnew - \fold) / \Delta t$,
but in the right-hand side we may use any combination of the old and the new function values.
However, even though the discretized Fokker--Planck equation conserves the mass exactly (in the absense of fluxes through boundaries), it does not automatically conserve energy. Consider the change in total energy $\scE$ in one timestep:
\begin{align*}
\Delta \scE \equiv
\int_0^\infty \big(f^\mathrm{new}(h) - f^\mathrm{old}(h)\big)\,E(h)\,\D h =
-\Delta t \int_0^\infty \frac{\d \mathcal F(h)}{\d h}\,E(h)\,\D h =
-\Delta t \big[\mathcal F(h)\,E(h)\big]\Big|_0^\infty + \Delta t \int_0^\infty \frac{\mathcal F(h)}{g(h)}\,\D h .
\end{align*}
For simplicity, we consider a single-component system with zero-flux boundary conditions.
In the numerical approximation of the flux $\mathcal F$ we may use a certain linear combination $f^\mathrm{E}$ of the old and new DF values for the evolving function, and another combination $f^\mathrm{R}$ to compute the relaxation matrix $\mathsf R$ (note that since the advection/diffusion coefficients linearly depend on the DF, so does the relaxation matrix). Using (\ref{eq:I_0}-\ref{eq:K_h},\ref{eq:FokkerPlanckCoefs}), we get
\begin{align*}
\frac{\Delta \scE}{\Delta t} =
-\int_0^\infty \D h \Bigg[\frac{f^\mathrm{E}(h)}{g(h)} \int_0^h \D h'\,f^\mathrm{R}(h') \Bigg]
-\int_0^\infty \D h \Bigg[\frac{\d f^\mathrm{E}(h)}{\d h}
 \int_0^\infty \D h' \frac{f^\mathrm{R}(h')\,\mathrm{min}(h,h')}{g(h')} \Bigg].
\end{align*}
We integrate the second term by parts and again note that the boundary term is zero.
\begin{align*}
\frac{\Delta \scE}{\Delta t} =
-\int_0^\infty \D h \int_0^h      \D h'\; \frac{f^\mathrm{E}(h)\,f^\mathrm{R}(h')}{g(h)} 
+\int_0^\infty \D h \int_h^\infty \D h'\; \frac{f^\mathrm{E}(h)\,f^\mathrm{R}(h')}{g(h')} .
\end{align*}
Finally we exchange the order of integration in the second term 
and then switch $h$ and $h'$:
\begin{align}  \label{eq:EnergyError}
\frac{\Delta \mathcal E}{\Delta t} = \int_0^\infty \D h \int_0^h \D h'\; 
\frac{f^\mathrm{E}(h')\,f^\mathrm{R}(h) - f^\mathrm{E}(h)\,f^\mathrm{R}(h')}{g(h)} .
\end{align}

To cancel the energy error, we may take $f^\mathrm{R}=f^\mathrm{E}$. If both are equal to $f^\mathrm{old}$, this corresponds to the explicit Euler method, which is however only conditionally stable and in practice would require very short timesteps. \citet{Epperlein1994} suggested an elegant linearization for an implicit Euler method: replace the r.h.s. of equation (\ref{eq:FokkerPlanckMatrix}) with $\mathsf{R}^\mathrm{old}\,\boldsymbol{f}^\mathrm{new} + \mathsf{R}^\mathrm{new}\,\boldsymbol{f}^\mathrm{old} - \mathsf{R}^\mathrm{old}\,\boldsymbol{f}^\mathrm{old}$. The energy error resulting from the first and the second term thus cancels because $f^\mathrm{E}$ and $f^\mathrm{R}$ are exchanged in equation (\ref{eq:EnergyError}), and the third term does not introduce any error. Since $\mathsf{R}^\mathrm{new}$ depends linearly on $\boldsymbol{f}^\mathrm{new}$, this results in an ordinary linear equation system to be solved at each step; however, its matrix is dense, not band-diagonal. We instead opt to retain the band-diagonal structure of the system by replacing $\mathsf{R}^\mathrm{new}$ with a linear extrapolation constructed from the previous timestep, and solve for the vector $\boldsymbol{f}$ only. We also experimented with a Crank--Nicolson scheme having an equal-weight symmetric combination of old and new $\mathsf{R}$ and $\boldsymbol f$ in the r.h.s., but found it to be only marginally better in terms of energy conservation, and prone to instability.

The joint evolution of the DF and the potential is followed using the operator-splitting approach: first we advance the DF for a timestep $\Delta t$ using the Fokker--Planck equation in a fixed potential $\Phi$, and then recompute the stellar density (\ref{eq:rho_from_f}) and potential (\ref{eq:Poisson}) in the Poisson step, while keeping $f(h)$ fixed. Since the potential $\Phi$ and the mapping between $\Phi$ and $h$ also enter the integral for the density, most previous studies updated it in several iterations. Instead we predict the potential $\tilde\Phi$ at the end of the timestep by linearly extrapolating its evolution from the previous timestep, recompute the density using this predicted $\tilde \Phi$ and its associated mapping between $E$ and $h$, and then use this density to update the potential. We have checked that this procedure is sufficiently accurate so that further iterations do not significantly improve it, and we perform it after each Fokker--Planck step. As shown in the appendix of \citet{Cohn1979}, the combination of Fokker--Planck and Poisson steps conserves the energy to within $\mathcal O(\Delta t^2)$ per timestep, and the error arising from performing only one iteration in our predictor/corrector scheme is also $\mathcal O(\Delta t^2)$.

We typically use a uniform grid in the scaled variable $x\equiv\ln h$ covering a sufficiently large range ($\gtrsim 20$ orders of magnitude) with a few hundred points (higher-order finite-element methods need fewer points), and extrapolate the DF outside the grid as a power-law in $h$. Other studies which used $f(E)$ typically employed some scaling transformations to increase the dynamical range and resolution, tailored to the specific problem, whereas in the case of $f(h)$ a uniform grid in $\ln h$ is always a reasonable choice: a significant change in the properties of the system (e.g., the slope of the density profile) is always accompanied by a significant change in $h$, even if it occurs in a relatively narrow range of $E$.
Density and potential are computed on a logarithmically-spaced grid in radius with $\sim 100$ nodes, which covers the extent of the grid in $h$, but needs not coincide with it. Various quantities such as $\Phi(r)$, $h(\Phi)$, $I_0(h)$ are represented by quintic splines, constructed from independently computed values and derivatives of the relevant function at grid nodes, which provide substantially higher interpolation accuracy than cubic splines (well below $10^{-8}$ with $\lesssim10$ nodes per decade).

\end{document}